\documentclass[10pt,sigplan,letterpaper,twocolumn]{article}
\usepackage[10pt,nocopyright]{sigmin}

\usepackage{tabularx}
\usepackage{inconsolata}
\usepackage{amssymb}
\usepackage{listings}
\usepackage{./listings-rust/listings-rust}
\usepackage{cite}
\usepackage{color}
\usepackage[hyphens]{url}                                         
\usepackage[breaklinks,colorlinks]{hyperref}                      
\usepackage[usenames,dvipsnames]{xcolor}                          
\hypersetup{citecolor=blue,linkcolor=blue}                        
\usepackage{amsmath,amsopn,amssymb,amsthm}                        
\usepackage{thmtools}
\usepackage{thmbox}
\usepackage[toc,page]{appendix}
\usepackage{subcaption}                                           
\usepackage{endnotes,microtype,xspace,fancyvrb,multirow} 
\usepackage{epigraph} 

\usepackage{booktabs}                                             
\usepackage{array,underscore,relsize}                             
\usepackage{libertine}
\usepackage[T1]{fontenc}                                          
\usepackage{times}                                                
\usepackage{fancyhdr,lastpage}                                    
\usepackage{enumitem}                                             
\usepackage[labelfont=bf,font=normal,skip=1pt]{caption}           
\usepackage[export]{adjustbox}                                    
\usepackage{breakurl}                                             
\usepackage{pifont}                                               
\usepackage{tablefootnote}                                        
\usepackage{bbding}                                        
\usepackage{longtable}

\usepackage[linesnumbered,vlined,boxed,commentsnumbered]{algorithm2e}
\DontPrintSemicolon
\newcommand{\ra}[1]{\renewcommand{\arraystretch}{#1}}

\usepackage{amssymb}
\usepackage[normalem]{ulem}
\usepackage[hang,flushmargin]{footmisc}
\usepackage{textcomp}
\usepackage{graphicx}
\usepackage{titling}

\usepackage[compact]{titlesec}
\titleformat*{\section}{\large\bfseries}
\titleformat*{\subsection}{\normalsize\bfseries}
\titleformat*{\subsubsection}{\normalsize\bfseries}
\titlespacing*{\section}{0pt}{*2}{*1}
\titlespacing*{\subsection}{0pt}{*1.5}{*0.8}

\hypersetup{
  colorlinks,
  linkcolor={red!50!black},
  citecolor={blue!50!black},
  urlcolor={blue!80!black},
}

\newcommand{\sys}{{\textsc{RepTX}}} 

\newcommand{\leaselock}{LLLock}
\newcommand{\rtx}{\emph{r}{TX}}

\def\eg{e.g.,~}

\newcommand{\stitle}[1]{\vspace{1.ex}\noindent{\bf #1}}

\newcommand{\noindentstitle}[1]{\noindent{\bf #1}}

\newcommand{\fig}[1]{Figure{~\ref{#1}}}
\newcommand{\evalline}[1]{\textbf{#1}}

\newcommand{\red}[1]{\textcolor[rgb]{1,0,0.4}{#1}}

\newcommand{\TODO}[1]{\textcolor{red}{TODO: #1}}

\usepackage{tikz}

\usepackage{balance}
\usepackage{authblk}

\newcommand{\race}{RaceHashing} 
\newcommand{\sherman}{Sherman}
\newcommand{\tree}{B$^+$Tree}
\newcommand{\drtmh}{DrTM$^+$H}

\begin{document}



\title{\Large \bf{Transactional Indexes on (RDMA or CXL-based) Disaggregated Memory with Repairable Transaction}}

\renewcommand\Authsep{, }
\renewcommand\Authand{, }

\setlength{\affilsep}{0.5em}
\author[1,2]{Xingda Wei\thanks{{Equal contribution.}}}
\author[1]{Haotian Wang\textsuperscript{*}}
\author[1]{Tianxia Wang}
\author[1,2]{Rong Chen}
\author[1]{Jinyu Gu}
\author[3]{Pengfei Zuo}
\author[1] {Haibo Chen}
\affil[1]{\vspace{-2.mm}Institute of Parallel and Distributed Systems, SEIEE, Shanghai Jiao Tong University\vspace{0.8mm}}
\affil[2]{Shanghai AI Laboratory\vspace{-1.mm}}
\affil[3]{Huawei Cloud\vspace{-1.mm}}

\date{}
\maketitle

\frenchspacing
\begin{abstract}    
    The failure atomic and isolated execution of clients operations     
    is a default requirement for a system that serve multiple loosely coupled clients at a server.
    However, disaggregated memory breaks this requirement in remote indexes
    because a client operation is disaggregated to multiple remote reads/writes.
    Current indexes focus on performance improvements and largely ignore tolerating client failures.

    We argue that a practical DM index should be \emph{transactional}: 
    each index operation should be failure atomic and isolated 
    in addition to being concurrency isolated. 
    We present repairable transaction ({\rtx}),
    a lightweight primitive to execute DM index operations.
    Each {\rtx} can detect other failed {\rtx}es on-the-fly 
    with the help of concurrency control.
    Upon detection, it will repair their non-atomic updates online with the help of logging, 
    thus hiding their failures from healthy clients.
    By further removing unnecessary logging and delegating concurrency control 
    to existing carefully-tuned index algorithms, 
    we show that transactional indexes can be built at a low performance overhead on disaggregated memory.
    We have refactored two state-of-the-art DM indexes, RaceHashing and Sherman ({\tree}), with {\rtx}. 
    Evaluations show that {\rtx} is 1.2--2$\times$
    faster than other alternatives, e.g., distributed transaction.
    Meanwhile,
    its overhead is up to 42\% compared to non-fault-tolerant indexes.

\end{abstract}    
\section{Introduction}
\label{sec:intro}

Disaggregated memory (DM), which splits the CPU and memory in monolithic servers 
into network-attached computing and memory pool, is getting prevalent 
in data centers~\cite{racehashing,DBLP:conf/osdi/GaoNKCH0RS16,themachine,
DBLP:conf/hotstorage/LegtchenkoWRDBD17, mind}.
Despite being better in elasticity and resource utilization, 
accessing the memory pool (memory node) from the computing pool (computing node) is much slower than accessing local memory, 
even with high-speed interconnects like RDMA or CXL~\cite{DBLP:conf/asplos/LiBHEZNSRLAHFB23, farm}.
Prior work built optimized remote indexes on disaggregated memory (DM indexes) for DM applications,
including hash tables~\cite{racehashing, DBLP:conf/sosp/WeiSCCC15},
{\tree}~\cite{DBLP:conf/sigmod/WangLS22,DBLP:conf/sigmod/0001VBFK19}, 
and many others~\cite{DBLP:conf/hotos/AguileraKNS19,rolex}, 
aiming to improve the application performance.
These indexes follow a client-server model where the index data is stored on the memory node (server),
and clients on the computing node access the index through the networked memory reads/writes.
Thus, they can serve as foundational building blocks for 
critical datacenter systems like distributed in-memory key-value stores (KVS)~\cite{racehashing,DBLP:conf/sigmod/WangLS22}.

\stitle{Problem.}
In traditional distributed systems like KVS, 
a server may simultaneously serve many loosely-coupled clients,
e.g., microservices~\cite{redis-microservice} or serverless functions~\cite{DBLP:conf/cidr/HellersteinFGSS19}, 
where clients are vulnerable to failures. 
Therefore, a basic (implicit) requirement is 
that each client operation should be all-or-nothing atomic under failures~\cite{cse-textbook} 
and failure isolated---a failed client cannot affect others (\textsection{\ref{sec:background-problem}}).
Traditional systems trivially achieve so because they use remote procedure calls (RPCs):
the client operation is shipped to the server for execution with a single RPC.
As long as the server is alive,
each operation is atomically executed at the server and the server is isolated from failed clients. 
However, DM decouples each client operation from a single network request 
to multiple networked DM reads/writes.
If a client fails before finishing all these writes, 
the index data on the server can become non-atomic (due to partial update) 
and unavailable to other clients (due to unreleased locks), although the server is alive.

\stitle{State-of-the-art DM indexes.}
Existing DM indexes~\cite{racehashing,DBLP:conf/sigmod/WangLS22, DBLP:conf/hotos/AguileraKNS19,rolex} 
focus on improving the index performance by optimizing the index structure layout to better fit DM
or supporting a fast concurrency control under limited DM primitives (e.g., memory read/write and atomics).
Despite performance improvements, 
none of them considered tolerating client failures, which is challenging 
due to the complexity of the indexes and non-atomic nature of DM index operations (\textsection{\ref{sec:background-existing-solutions}}).
Even for a single index, 
it is hard to derive a correct algorithm to tolerate non-atomic updates 
at the client-side.
Take {\tree} as an example.
A failed client insertion can leave the index data in many intermediate states.
For failure atomicity and isolation,
other clients must repair all these cases on-the-fly,
since stopping all the clients for a recovery is unacceptable for a distributed system.
Worse even, it is impossible to distinguish a failed operation from an unfinished one.
Blindly repairing an (non-failed) unfinished operation leads to data corruption.

\stitle{Goal: a transactional index.}
The necessities of failure atomicity and isolation,
plus the fact that a \emph{consistent} index must provide \emph{concurrency isolation} between clients,
motivate our argument that a DM index should be \emph{transactional}: 
a DM index should provide failure atomic, failure isolated and concurrency isolated index operations.

A straightforward solution, therefore, 
is to build indexes with distributed transactions for DM~\cite{farm, farmv2, drtmh, DBLP:conf/sosp/WeiSCCC15, ford-fast22, DBLP:conf/eurosys/ChenWSCC16}.
Unfortunately, techniques from transactions are too heavyweight for DM indexes 
because they don't consider the setup where clients are loosely coupled. 
Specifically, they need explicitly coordination between clients upon detecting client failures, 
which not only breaks the client failure isolation,
but also consumes extra system resources and adding deployment burden to DM indexes
(\textsection{\ref{sec:background-existing-solutions}}).


\stitle{Our approach: \emph{repairable transaction} ({\rtx}).}
We propose {\rtx}, a lightweight transaction primitive to build transactional DM indexes. 
A {\rtx} can detect other failed {\rtx}es during its reads/writes to DM indexes.
More importantly,
upon detection, it will instantly repair any failure states by helping 
the failed {\rtx} to commit or abort. 
We achieve so without breaking the loosely-coupled model of DM index applications:
each {\rtx} is fully implemented with DM primitives
and does not communicate with other clients.

{\rtx} achieves online index repair with the help of write-ahead logging, 
the de facto technique for failure atomicity.
Therefore, 
it can automatically repair the failed index data no matter how complex the indexes are. 
Nevertheless, we still need to work with the unique constraints of DM indexes (\textsection{\ref{sec:overview-challenges}}),
i.e., the lack of computing power at the memory node~\cite{racehashing, 
DBLP:conf/osdi/GaoNKCH0RS16,themachine, DBLP:conf/hotstorage/LegtchenkoWRDBD17,
DBLP:conf/asplos/LiBHEZNSRLAHFB23} and loosely coupled clients. 
First, 
how can one {\rtx} effectively detect another's failure without communicating to it?  
The memory node's weak computing power can't support failure detection.
Second, given that the repaired {\rtx} may be still committing, 
how can the repairing {\rtx} correctly help it to commit without coordinating with it?

To efficiently and effectively detect failed {\rtx}es,
our insight is that failed execution of concurrency control will leave hints on the index metadata, 
e.g., a lock that has not been released for a period.
Therefore, 
we propose a new lock primitive {\leaselock} (\textsection{\ref{sec:lease-lock}}),
which supports piggybacking failure detection with normal lock acquire operation 
on existing DM locks.
To correctly coordinate concurrent commit of the repairing and repaired {\rtx}s,
we further design a new \emph{idempotent} commit protocol (\textsection{\ref{sec:tx}}).
The observation is that as long as the commit is idempotent,
different clients can correctly commit the same {\rtx} without coordination.

{\rtx} relies on the concurrency control protocol for failure detection and correct index repair.
Therefore, 
we incorporate optimistic concurrency control (OCC)~\cite{DBLP:journals/tods/KungR81}
to isolate concurrent index operations by default.
A natural question is that whether a general concurrency control protocol like OCC 
as well as logging will bring significant performance overhead to DM indexes?
We show it is possible (\textsection{\ref{sec:opt-impl}}) to minimize the overhead by 
(1) designing an optimized protocol to remove logging for index operation that only modifies one object,
which is common in DM indexes
 and (2) delegating part of the concurrency control to existing algorithm with smarter conflict detection, 
 inspired by existing work~\cite{DBLP:conf/eurosys/HermanIHTKLS16}.
Our results on YCSB benchmark~\cite{ycsb} demonstrated that 
the cost of {\rtx} is 
up to 42\% compared to existing non-atomic indexes. 

{\rtx} can be used to build complex DM indexes from scratch,
e.g., in {\tree} where it is challenging to derive a transactional index algorithm. 
On the other hand, 
we can also use it to augment existing indexes where most operations is transactional.
For example, {\race} supports transactional operations as long as no rehashing happens. 
We extend it with {\rtx} to support transactional rehashing while bring minimal overhead
to its normal operations.

We have prototyped a runtime {\sys} that implements {\rtx}. 
 To demonstrate the efficiency and efficacy, 
 we have refactored two state-of-the-art DM indexes using {\sys}: 
 RaceHashing~\cite{racehashing} (HashTable) and Sherman~\cite{DBLP:conf/sigmod/WangLS22} (B$^+$Tree). 
 Extensive evaluations using industry-standard YCSB benchmark~\cite{ycsb} show that 
 {\sys} is 1.2--2$\times$ faster than existing alternatives for building DM indexes, 
 e.g., using distributed transactions on DM (i.e., {\drtmh}~\cite{drtmh}).
 Besides, {\sys} can repair a crashed client with no observable performance collapse in the interval of 1\,ms. 

 \stitle{Contributions}. We highlight our contributions as follows: \\[-18pt]
\begin{itemize}[leftmargin=*,leftmargin=10pt,itemindent=0pt]
    \item \textbf{Problem and transactional DM index}: We show the necessity of 
    tolerating client-side failures for DM indexes.
    We characterize property of practical DM indexes as transactional: 
    each index operation should be failure atomic, failure isolated and concurrency isolated. \\[-18pt]
    
    \item \textbf{Repairable transaction for transactional DM indexes}:
    We propose using repairable transaction (\rtx) to build transactional DM indexes. 
    {\rtx} is both efficient and lightweight for DM indexes. \\[-18pt]

    \item \textbf{Demonstration}:
    An implementation of the repairable transaction in {\sys}, along with a refactor of existing, 
    carefully-tuned DM indexes, demonstrates the feasibility and efficiency of our approach.
\end{itemize}

\section{Background and Motivation}
\label{sec:background}

\begin{figure}[!t]
    \vspace{-2mm}
    \begin{minipage}{1\linewidth}
    \hspace{-3mm}
    \centering    
    \includegraphics[scale=1.1]{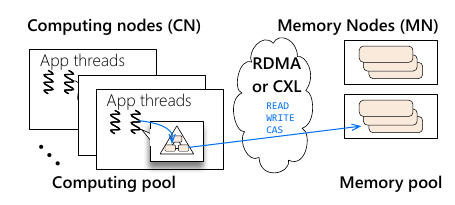}
    \end{minipage} \\[5pt]
    \begin{minipage}{1\linewidth}
    \caption{\small{%
    An overview of disaggregated memory architecture.
    }}
    \label{fig:setup}
    \end{minipage} \\[-20pt]
    \end{figure}

\subsection{Disaggregated memory (DM) and DM indexes}
\label{sec:background-dm}

\noindentstitle{Disaggregated memory.}
Servers of disaggregated memory architecture~\cite{racehashing, 
DBLP:conf/osdi/GaoNKCH0RS16,themachine, DBLP:conf/hotstorage/LegtchenkoWRDBD17,
DBLP:conf/asplos/LiBHEZNSRLAHFB23}
are separated into two groups, namely computing and memory pool. 
As shown in {\fig{fig:setup}}, 
computing pool hosts nodes with powerful CPUs, but with limited memory.
Memory pool has nodes with huge memory, but with weak CPUs (or no CPU at all). 
Without losing generality, we term the nodes in the computing pool as \emph{computing node}s (CNs) 
and nodes in the memory pool as \emph{memory node}s (MNs). 
The advantage of disaggregation is that resources in different pools 
can be scaled and fault-tolerant independently, 
improving resource utilization and elasticity 
of the applications~\cite{firebox,DBLP:conf/osdi/GaoNKCH0RS16, DBLP:conf/isca/LimCMRRW09, DBLP:conf/osdi/WangMLLRNBNKX20, racehashing, DBLP:conf/hpca/LimTSACRW12}. 

Under DM, applications run on the CNs and access data stored on the MNs.  
For efficiency, CNs and MNs are bridged via high-speed interconnects 
like RDMA or CXL~\cite{DBLP:conf/usenix/GoukLKJ22}.\footnote{\footnotesize{CXL 3.0 supports 
memory sharing between CNs~\cite{cxl}.}}
From a CN's perspective, the MN memory provides an asynchronous shared memory access model,
where applications can use the \emph{DM primitives} to \texttt{READ}, \texttt{WRITE}, 
\texttt{CAS} (compare-and-swap), and \texttt{FAA} (fetch-and-add) the MN memory.
Like DRAM, the atomicity of the DM primitives is limited.
For example, RDMA only provides 64\,B atomic \texttt{READ}/\texttt{WRITE}~\cite{farm,DBLP:conf/eurosys/ChenWSCC16}
while \texttt{ATOMIC}s only supports 8\,B.

\stitle{DM indexes.}
It is challenging to utilize DM: 
even with high-speed interconnects, 
accessing data on the MN is slower than the local memory of CN~\cite{DBLP:conf/asplos/LiBHEZNSRLAHFB23,farm}.
Therefore, prior work builds DM indexes including HashTable~\cite{racehashing,DBLP:conf/sosp/WeiSCCC15} 
and {\tree}~\cite{DBLP:conf/sigmod/WangLS22,DBLP:conf/sigmod/0001VBFK19}.
These indexes are highly optimized for performance with 
careful data structure organization and index access algorithms.
Despite complex internal design, they provide a simple yet powerful developer-familiar interface, 
i.e., \texttt{Get} and \texttt{Insert}, 
so it is easy to utilize them in distributed systems like 
distributed key-value store~\cite{racehashing,DBLP:conf/sosp/WeiSCCC15}.

\subsection{Motivation of transactional DM indexes}
\label{sec:background-problem}

\begin{figure}
    \begin{minipage}{1\linewidth}
    \hspace{-3mm}
    \centering    
    \includegraphics[scale=1.1]{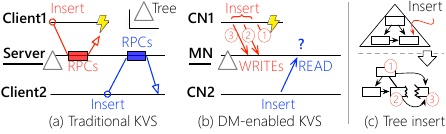}
    \end{minipage} \\[8pt]
    \begin{minipage}{1\linewidth}
    \caption{\small{%
        An overview of distributed key-value store (KVS)
        using (a) RPCs and (b) DM primitives. 
        (c) An illustration of tree insertion.
    }}
    \label{fig:redis}
    \end{minipage} \\[-20pt]
\end{figure}

\noindentstitle{A motivating example: distributed key-value store (KVS).}
We motivate the necessity of failure atomic and isolated DM indexes by presenting the workflow of KVS, 
a typical use case of DM indexes. 
KVS (e.g., redis~\cite{redis}) consists of two components: 
(1) data structures stored at the server memory, 
containing key-value pairs and an index to these pairs, 
and 
(2) a library at the client, 
implementing index operations to get or insert key-value pairs on the server data.
For flexibility and elasticity,
the library can be linked to loosely-coupled applications running on different CNs,
e.g., microservices~\cite{redis-microservice} or serverless functions~\cite{krcore,DBLP:journals/pvldb/SreekantiWLSGHT20}. 

In traditional KVS ({\fig{fig:redis}} (a)), the CN executes the index operations with RPCs.
Under DM, 
the client replaces the server-side index with DM indexes and 
leverages DM primitives to execute operations (b).
For example, if the index is a {\tree}, 
the CN executes an \texttt{Insert} by 
issuing multiple \texttt{WRITE}s to update different nodes.
To coordinate concurrent writes,
CN also acquires locks using \texttt{CAS}es~\cite{racehashing,DBLP:conf/sigmod/WangLS22}.

\stitle{Client failures in KVS.}
In KVS applications, clients are loosely coupled so they 
do not directly communicate with or even know each other.
Consequently, providing failure isolation among clients is important: 
a client does not expect another client it has never heard of
before to block its access to the KVS.
Besides, failure isolation dramatically simplifies application development: 
they only need to take care of the interactions with the KVS,
not others who are simultaneously using the same KVS instance. 

It is trivial for existing RPC-based KVS to tolerate client failures.
For them, each client request is atomically processed by the server. 
Therefore, as long as the server is alive, 
a failed client will not affect the server's processing of other clients' request,
as shown in {\fig{fig:redis}} (a).
To the best of our knowledge, 
all existing RPC-based KVSes like Redis or memcached provide client failure isolation. 

On the other hand, tolerating client failures is challenging in DM indexes, 
because DM breaks the index operation into separated \texttt{WRITE}s.
As shown in {\fig{fig:redis}} (b) and (c),
an insertion on tree-based index may involve at least three writes for internal and leaf nodes
(\ding{192}--\ding{194}).
If CN1 crashes, the index will be left in an inconsistent state. 
Even worse, 
a failed CN can block other CNs from accessing the index (not shown in the figure). 
Specifically,
DM indexes all leverage \texttt{CAS}-based spinlock to coordinate concurrent accesses 
from different CNs~\cite{DBLP:conf/sigmod/WangLS22,DBLP:conf/sigmod/0001VBFK19,
DBLP:conf/sosp/WeiSCCC15,racehashing}.
Even for {\race}~\cite{racehashing} that is designed for lock-free operations, 
it still needs to acquire lock under rehashing. 
If the lock holder fails before releasing the lock, 
future CNs will be blocked without progress, even the updates to the index are atomic. 

\stitle{Goal: failure atomicity and failure isolation.}
We argue that a practical DM index should be resilient to client failures. 
Specifically, its operations should satisfy the following two properties: 
(1) \emph{Failure atomicity}.
Each index operation should happen completely or not at all on the MN even its executing client fails
(a.k.a, all-or-nothing atomicity~\cite{cse-textbook}).
(2) \emph{Failure isolation}.
The partially executed operation at a failed client 
should be isolated from other clients.

\subsection{Challenges for failure atomic and isolated indexes}
\label{sec:background-existing-solutions}

\noindentstitle{Correctly repairing complex failure states.}
First, a failed client can leave many inconsistent states on MN's index data.  
As shown in {\fig{fig:redis}} (c), 
if an insertion causes a split on a {\tree},
the data would be left in a maximum of six possible inconsistent states.
To isolate a client from failed client, 
we must quickly repair the index to a consistent state as well as tolerate inconsistent data read during its execution.
Though enforcing a strict write order
(e.g., first write the leaves than the internal nodes~\cite{DBLP:journals/dr/Shasha99d})
can reduce the failed states, 
the detailed method is highly dependent on the 
index structure and their access algorithms. 

\begin{figure*}
    \begin{minipage}{1\linewidth}
        \centering    
        \includegraphics[scale=1.06]{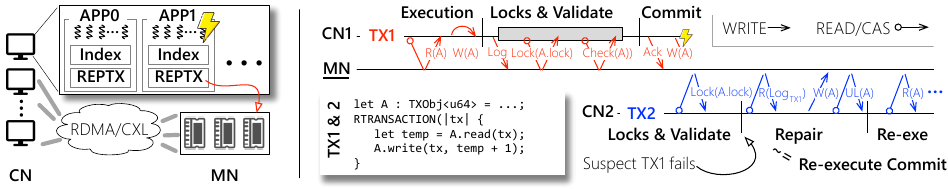} \\[5pt]
    \end{minipage}  \\[5pt]
    \begin{minipage}{1\linewidth}
    \caption{\small{%
    (a) System model of {\sys}.
    (b) {\rtx} execution flow.
    \textbf{W}: Write. \textbf{R}: Read. \textbf{Log}: Write-ahead logging. \textbf{UL}: Unlock.
    }}
    \label{fig:combined}
    \end{minipage} \\[-20pt]
\end{figure*}

\stitle{Effectively detecting failures.}
Even we derive a correct algorithm to repair all the inconsistent states, 
how to determine when an operation fails so that we can repair is another challenge.
An inconsistent state does not always imply a failed client:
it may have not finished its operation. 
A classic way to detect failure is through heartbeats.
However, MNs have weak or no CPU for doing so. 
On the other hand, 
sending heartbeats between CNs breaks the loosely-coupled assumption of clients 
and is also inefficient.
Leveraging a replicated state machine---a common technique 
adopted by distributed transactions on DM~\cite{farm, farmv2, drtmh, DBLP:conf/sosp/WeiSCCC15, ford-fast22, DBLP:conf/eurosys/ChenWSCC16}---for the heartbeats 
is possible but is usually too heavy for DM indexes (see the next challenge).

\stitle{Correctly coordinating the index repair.}
Even with an effective failure detection method, 
it is intractable to determine whether a client truly fails 
or it is just slow in a distributed system like DM~\cite{DBLP:journals/jacm/FischerLP85}. 
As a result, 
naively repair a (suspected) failed operation can lead to an irrecoverable state.
For example, if the repair releases the original client's lock, 
race conditions can happen (see also \textsection{\ref{sec:lease-lock}}).

Existing system explicitly coordinates clients,
i.e., reject the suspected client before the repair can happen. 
These coordination is typically done with a global coordination service 
implemented with replicated state machine~\cite{farm,farmv2,drtmh,DBLP:conf/sosp/WeiSCCC15,ford-fast22,DBLP:conf/eurosys/ChenWSCC16,fuse-fast23}.
Unfortunately, such a coordination is too heavyweight for DM indexes. 
First, it breaks the failure isolation between clients (for rejection). 
Rejecting healthy clients is costly under DM:
it takes tens or hundreds of milliseconds for a client to reconnect~\cite{fuse-fast23,krcore} so we should carefully avoid it. 
Worse even, it is even unknown how to support the rejection in a pure hardware solution like CXL. 
Second, the coordination service pays for the extra resources and management overhead. 
In existing KVS (e.g., Redis), clients never enable such mechanism 
to tolerate failures of other clients.
\section{Overview}
\label{sec:overview}

\label{sec:system-overview}

\noindentstitle{System setup and failure model.}
As shown in {\fig{fig:combined} (a)},
our setup is the same as existing DM indexes: 
{\sys} is a library linked with applications running on the CNs to access indexes on the MN. 
One MN can be shared by multiple clients. 
We assume applications on the CNs can fail, but MNs rarely fail.
First, for a single node, MN is more robust in hardware,
because the failure probability of its core component (DRAM) 
is much smaller than CN CPU~\cite{DBLP:conf/eurosys/NightingaleDO11}.
Second, CNs are vulnerable to software failures while MNs are not.
According to large-scale empirical results,
the likelihood of machine failures caused by software errors 
is orders of magnitude higher than hardware ones~\cite{dcascomputer}.
Finally, one MN is shared by multiple clients, 
so the likelihood of client failure aggregates. 
Note that we do not consider Byzantine failures 
and treat broken network as failed CNs without losing generality.

\stitle{Approach: repairable transaction (\rtx).}
To relieve developers from designing complex algorithms for tolerating failed index operations,
we propose a primitive called {\rtx}. 
The developers can use it to build DM indexes by
(1) wrap data object related to the index with a \texttt{TXObj},
(2) using {\rtx} read/write to implement the index algorithm atop of the wrapped objects and 
(3) wrapping reads/writes with a \texttt{RTRANSACTION} block, see below: 
\begin{lstlisting}[language=rust,label=rust, 
    lineskip=1pt, 
    xleftmargin=0pt,xrightmargin=0pt]
let r : Ptr<TXObj<Node>> = ...; // Tree root node
let res = RTRANSACTION(|tx| { // Insertion w/ rTX
    let root_value = tx.read(r);       // DM read
    for child in root_value {     
        ...                 // Detailed algorithm
} });                              // rTX commits
\end{lstlisting}
{\rtx} runtime handles failure atomicity, isolation and concurrency control.

\stitle{{\rtx} execution flow.}
{\rtx} uses logging for failure atomicity, 
optimistic concurrency control~\cite{DBLP:journals/tods/KungR81} (OCC) for isolating concurrent accesses
and an online repair protocol for failure isolation.
{\fig{fig:combined}} (b) presents the execution flow of how an {\rtx} (TX2) 
repairs a failed {\rtx} (TX1).
TX1 first executes the index operation with an \emph{Execution} phase:
it reads the data pointed via \texttt{READ}\footnote{\footnotesize{
    If the value has been read or updated by the {\rtx} before,
    {\sys} will return the local copy instead of re-reading the data.
}} 
and writes the update to a local copy at its writeset.\footnote{\footnotesize{If the CN memory is insufficient, 
we will swap the local write to the MN as it will eventually write to MN upon commits.}}
Afterward, TX1 enters the \emph{Validation} phase, 
where all data in the writeset are locked and readset validated,
i.e., whether it has changed after the first read in the execution phase. 
The locks and validates are implemented using \texttt{READ}s and \texttt{CAS}es.
Before the validation, TX1 writes a tentative log to the MN for future repair.
If the validation passes,
the TX will \emph{commit} by transferring the log to a committed state.
Finally, TX1 will update the data and release the locks.

TX2 will detect TX1's failure by accessing a data modified by it.
Afterward, TX2 enters a \emph{Repair} phase, 
which will read TX1's log and re-execute its commit using the log,
i.e., updating its data releasing its locks. 

\subsection{Challenges and overview of solutions}
\label{sec:overview-challenges}

With logging, 
{\sys} can correctly repair the index under complex failure states
(\textsection{\ref{sec:background-existing-solutions}}).
The remaining challenges are: \\[-18pt]
\begin{enumerate}[leftmargin=*,leftmargin=10pt,itemindent=0pt]
    \item \emph{Effective failure detection.} 
    How can one {\rtx} quickly detect another failure? 

    \item \emph{Lightweight and correct repair coordination.}
    An {\rtx} (TX2) repairs another (TX1) by re-executing TX1's commit phase. 
    However, TX1 may be still committing (falsely suspected). 
    Without coordination---let them concurrently updating the same writeset and release the same set of locks---result 
    in race conditions (see \textsection{\ref{sec:lease-lock}}). 
\end{enumerate} 

\stitle{Overview of solutions.}
For the first challenge, 
our insight is that the concurrency control protocol gives us hints on whether an {\rtx} fails.
Specifically, in a common concurrency control protocol,
a failed {\rtx} that breaks failure atomicity must have multiple updates, 
so it must leave some data in a locked state if failed.
Therefore, if an {\rtx} finds another that has locked a data for a long time,
we suspect it as failed.
Hence, 
we propose a new lock primitive called {\leaselock} (see \textsection{\ref{sec:lease-lock}})
that supports timeout detection and log owner tracing.

For the second challenge, 
we find coordination is not necessary needed as long as the re-executions of the commit phase 
behave as they were executed only once (i.e., idempotence). 
More importantly, idempotent OCC commit can be achieved in DM by \emph{using} \texttt{CAS}es \emph{to update values and release locks}. 
Specifically, we separate the data layout (\textsection{\ref{sec:data-structure}}) to allow a copy-on-write update scheme. 
Afterward, 
we atomically update the data with \texttt{CAS}. 
Since the \texttt{CAS} can only succeed once, 
the update is idempotent under re-execution (\textsection{\ref{sec:tx}})
and won't affect future modifications.

\section{Design and Implementation}
\label{sec:design}


\subsection{Logged lease lock (\leaselock)}
\label{sec:lease-lock}

\begin{figure}[!t]
    \begin{minipage}{1\linewidth}
    \vspace{5pt}
    \hspace{-6mm}
    \centering    
    \includegraphics[scale=0.93]{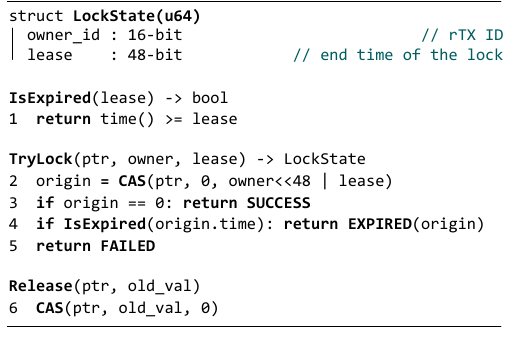}
    \end{minipage} \\[-8pt]
    \begin{minipage}{1\linewidth}
    \caption{\small{%
    The pseudo-code of lease lock executed at the CN. 
    }}
    \label{fig:alg-lease-lock}
    \end{minipage} \\[-20pt]
    \end{figure}

We use {\leaselock} to coordinate shared accesses 
and detect {\rtx} failures.
Normally, it can be treated as a spinlock.

\stitle{Basic scheme.} 
Each {\leaselock} occupies 8B of memory (represented as integer) on the MN. 
The integer encodes the following information,
as shown in {\fig{fig:alg-lease-lock}} (\texttt{LockState}): 
(1) a \emph{lease} that grants the lock to the holder until the lease time
and (2) the {\rtx} ID of the current lock holder. 
To allow direct comparison between different clients, 
we use POSIX time as the lease.
Since the number of bit available in the LockState (only 48 bits) is limited, 
we restrict the lease precision to milliseconds, 
which we found is sufficient for a quick repair
 (see also \textsection{\ref{sec:eval-lease}}).
{\sys} uses atomic \texttt{CAS} operations to acquire and release the lock, 
similar to a spinlock (Line 2--6).

\stitle{The false release problem and solution.}
The expiration of {\leaselock} does not necessary means a failed holder (Line 1). 
If the holder is still processing, 
releasing {\leaselock} would cause \emph{false release}, 
because two {\rtx}s will be concurrently updating the same data,
both believing that they hold the lock.

\begin{figure}[!t]
    \begin{minipage}{1\linewidth}
    \hspace{-3mm}
    \centering    
    \includegraphics[scale=1.05]{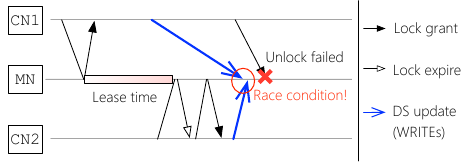}
    \end{minipage} \\[5pt]
    \begin{minipage}{1\linewidth}
    \caption{\small{%
    An illustration of the false release caused by ``fly WRITE'', 
    even all CN clocks are the same.
    }}
    \label{fig:lease-lock-problem}
    \end{minipage} \\[-15pt]
    \end{figure}

Several factors can cause false releases.
A straightforward one is unsynchronized clock---the time of different CNs drift. 
Although synchronized clocks (e.g., TrueTime~\cite{DBLP:conf/osdi/CorbettDEFFFGGHHHKKLLMMNQRRSSTWW12, farmv2}) may alleviate the issue,
they are insufficient due to another phenomenon we called ``fly WRITE''.
Specifically, a WRITE ``flies'' if it is delayed due to the asynchronous communication nature of DM.
When the delay occurs, we cannot guarantee that the WRITE protected by a {\leaselock} finishes within its lease,
so the final result is the same as if the data was released before the WRITE finishes. 
{\fig{fig:lease-lock-problem}} illustrates this case, assuming all CNs' clocks are the same.

A naive solution is to reject accesses from suspected CN~\cite{dragojevic2015nocompromises, DBLP:conf/eurosys/ChenWSCC16}.
Unfortunately, rejecting falsely suspected CNs is costly and breaks the failure isolation between clients (see \textsection{\ref{sec:background-existing-solutions}}).
{\sys} instead tolerates the false release by considering it in the {\rtx} execution and repair protocol.
Specifically, 
we delay the {\leaselock} release until we can ensure the non-finished operations are done, 
and a done operation cannot overwrite any future requests. 
At a high level, to release an expired lock, 
we first inspect the execution state of its holder using the \texttt{owner\_id} in the {\leaselock}.
If the log indicates the holder is unfinished, 
we will re-execute the idempotent commit protocol to finish its writes and release the lock.
\textsection{\ref{sec:tx}} describes our protocols in detail.

\subsection{Object, log and memory organization at DM}
\label{sec:data-structure}

\vspace{-12pt}
\begin{figure}[!h]
    \begin{minipage}{1\linewidth}
    \hspace{-5mm}
    \centering    
    \includegraphics[scale=1.0]{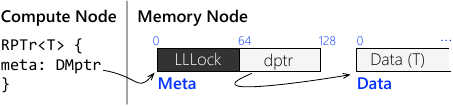}
    \end{minipage} \\[5pt]

    \end{figure}

\vspace{-20pt}
\stitle{\texttt{TXObj}.}
{\rtx} tracks conflicts with \emph{TXObj}s.
It is a wrapper over the index object (e.g., tree node), 
encoding necessary metadata for {\rtx}.
As shown in the above figure, 
each \emph{TXObj} contains a {\leaselock} and a pointer to the actual index data. 
The object is uniquely identified by the metadata pointer,
which is recorded in a \texttt{TXObj<T>} struct at the CN-side.

Our layout differs from the traditional approach that packs the data with the metadata.
The design aims at providing idempotent commit:
when updating the value, we can use \texttt{CAS} to install it atomically.
Meanwhile, it can reduce the log size to repair (see the below),
because the {\rtx} doesn't need to store the complete updated values.
The drawback is that reads the value requires an additional \texttt{READ} due to pointer chasing.
It is negligible for index with large objects (e.g., {\tree} node),
but should be used carefully for small objects. 
Note that updating the data does not modify the \texttt{TXObj<T>} 
since metadata pointer is left unchanged.
\vspace{-5pt}

\begin{figure}[!h]
    \begin{minipage}{1\linewidth}
    \hspace{-1mm}
    \centering    
    \includegraphics[scale=1.0]{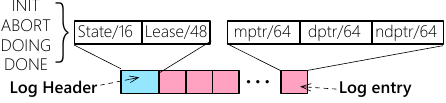}
    \end{minipage} \\[5pt]
    \end{figure}    

\vspace{-18pt}
\stitle{{\rtx} log.}
Each {\rtx} maintains a log at the MN whose layout is shown above.
\emph{INIT}---{\rtx} is in the execution phase, 
\emph{ABORT}---{\rtx} is aborted, with locks unreleased,
\emph{DOING}---{\rtx} is in the commit phase,
and 
\emph{DONE}---{\rtx} is done (either committed or aborted) and no further action is required.

If lease expires,
other {\rtx} can read the log from {\leaselock} to repair the index. 
To avoid redundant repair and tolerate failures during the repair, 
we encode a repair lease at the log header.
Different repair (and executing) can use the this lease to coordinate whether they need to repair.
Note that the lease is only valid when the log is in the \emph{DOING} state.

The remaining of the log (log entries) records the writeset of the TX. 
Thanks to a separate object layout, 
each entry is extremely compact:
it only records the object pointer (\texttt{mptr}),
the old data (\texttt{dptr}) and the updated pointer (\texttt{ndptr}),
instead of the whole value (e.g., {\tree} node).

\begin{figure}[!t]
    \begin{minipage}{1\linewidth}
    \vspace{-3mm}
    \hspace{-5mm}
    \centering    
    \includegraphics[scale=1.1]{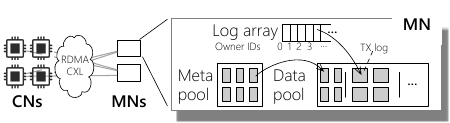}
    \end{minipage} \\[5pt]
    \begin{minipage}{1\linewidth}
    \caption{\small{%
        Memory organization in {\sys} at the DM.
    }}
    \label{fig:memory-layout}
    \end{minipage} \\[-20pt]
    \end{figure}

\stitle{Overall MN memory organization.}
We organize the MN memory with the following three parts ({\fig{fig:memory-layout}}): 
a log array, a meta pool and a data pool.
The log array is used to find the TX's log buffer given the {\rtx} ID (\texttt{owner\_id} in the {\leaselock}).
Note that we reuse log buffers for {\rtx}s executed in a single thread, 
so in the common case {\rtx} does not modify the log array (see \textsection{\ref{sec:resource}}). 

\subsection{{\rtx} execution, commit and repair}
\label{sec:tx}

\begin{figure}[!t]
    \begin{minipage}{1\linewidth}
    \hspace{-3mm}
    \includegraphics[scale=0.93]{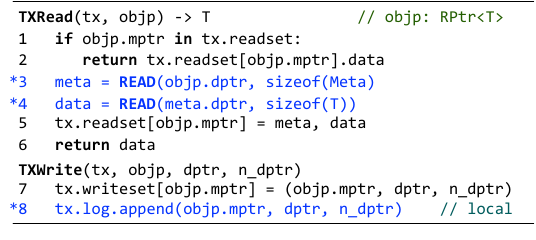}
    \end{minipage} \\[1pt]
    \begin{minipage}{1\linewidth}
    \caption{\small{%
    The simplified pseudo-code of reads/writes of {\rtx}.
    Lines marked as * are extension for failure atomicity and isolation.
    }}
    \label{fig:occ-readwrite}
    \end{minipage} \\[-15pt]
    \end{figure}

\noindentstitle{Assumptions in this section.}
For ease of presentation,
we first assume 
each {\rtx} will acquire a unique ID and log buffer
that has never been used by others before.
Moreover, when updating an \texttt{TXObj}, 
we will allocate a new data pointer and never reuse the old pointers.
\textsection{\ref{sec:resource}} removes these assumptions.

\stitle{The execution phase. } 
On entering the \texttt{RTRANSACTION} block,
{\sys} initializes an empty readset and writeset at the calling thread.
For read, if the pointer has never been read by the TX,
{\sys} issues two \texttt{READ}s to fetch the latest data from MN 
and records the data in the readset (Line 3--5 in {\fig{fig:occ-readwrite}}). 
Otherwise, 
{\sys} directly returns the previous copy. 
For write, 
we store the updated value and its pointer 
at the local writeset and local log, respectively (Line 7--8).

\begin{figure}[!t]
    \begin{minipage}{1\linewidth}
    \hspace{-7pt}
    \includegraphics[scale=0.93]{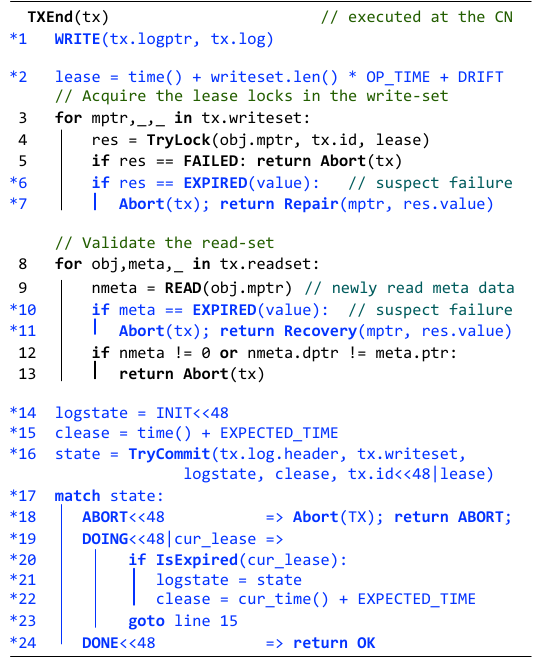}
    \end{minipage} \\[8pt]
    \begin{minipage}{1\linewidth}
    \caption{\small{%
    The pseudo-code of commit protocol of {\sys}. 
    Lines marked as * are extension to support failure atomicity and isolation.
    }}
    \label{fig:occ}
    \end{minipage} \\[-20pt]
    \end{figure}    

At the end of the \texttt{RTRANSACTION} block, 
{\sys} executes the protocol listed in {\fig{fig:occ}} to commit an {\rtx}.
The protocol contains the following phases: 

\stitle{The lock and validation phase (Line 1--13). }
First, 
we write the current log to the MN for future repair (Line 1). 
Then, we lock all the objects in the writeset via {\leaselock}s. 
Before locking, we calculate the lease of the locks and 
use the same lease for all the objects. 
This simplifies the logging and repair, 
e.g., we don't need to record the leases in the log. 

A good lease should be short for fast detection.
However, 
it cannot be arbitrarily short because it will cause false suspects. 
We regulate the lease based on the estimated executing time of \texttt{TXEnd} 
plus a constant \texttt{DRIFT} to tolerate unsynchronized CN clocks (Line 2) .
On our platform, we found 8\,ms is a good value (see \textsection{\ref{sec:eval-lease}}).

If all lock acquisitions succeed, 
{\sys} moves on to validate whether the reads 
have been modified since the execution phase. 
If no value has been modified, the {\rtx} can commit. 

\stitle{Commit (Line 14--24). }
The {\rtx} calls the \texttt{TryCommit} to commit. 
The call can fail---because another {\rtx} has been trying to commit it.
For such cases, we will retry until 
the invocation reports commit or abort (Line 17--24).

\begin{figure}[!t]
    \begin{minipage}{1\linewidth}
    \hspace{-5pt}
    \includegraphics[scale=0.93]{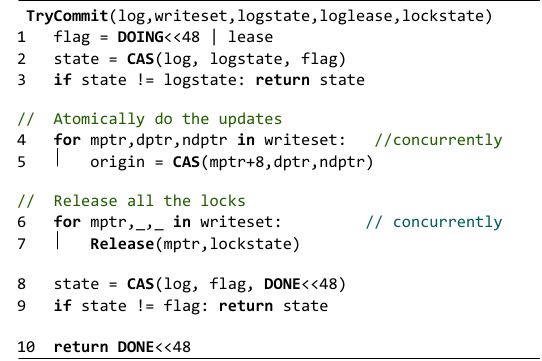}
    \end{minipage} \\[8pt]
    \begin{minipage}{1\linewidth}
    \caption{\small{%
    The simplified pseudo-code of \texttt{TryCommit}. 
    }}
    \label{fig:occ-trycommit}
    \end{minipage} \\[-25pt]
    \end{figure}

The \texttt{TryCommit} ({\fig{fig:occ-trycommit}})  realizes an idempotent commit phase, 
i.e., re-execution of it would result in the same effect as it were executed only once with no failure. 
To prevent unnecessary re-executions, 
it proceed if and only if we can atomically transfer the log into a \emph{DOING} state (Line 1--3).
If the transformation fails, e.g., the log is not in an \emph{INIT} state, 
then either another CN is committing the {\rtx} due to expired lease lock
(the log is in \emph{DOING} state with a valid lease), or 
the {\rtx} has committed (\emph{DONE}) or aborted (\emph{ABORT}).
None of the above cases require further execution.

We ensure the idempotence of \texttt{TryCommit} by leveraging the 
\texttt{CAS}es to  
update values in the writeset (Line 4--5) and release locks (Line 8--9).
If the value has been updated (or unlocked) by another \texttt{TryCommit},
the \texttt{CAS} will leave these values unchanged (Line 5 and 7). 
After releasing all the locks, we turn the log into the \emph{DONE} state (Line 8). 
Note that the state change must be done after the locks are released.
Otherwise, 
a committed {\rtx} may leave unreleased locks that needed to be released via retrying \texttt{TryCommit}.

\stitle{The repair phase. }    
When validating an {\rtx}, 
we will check the lease states of the locks 
(Line 5--6 and 11--12 in {\fig{fig:occ}}) to detect possible failed {\rtx}s.
If the lease expires, the executing {\rtx} aborts and repair the suspected operation.

\begin{figure}[!t]
    \begin{minipage}{1\linewidth}
    \hspace{-11pt}
    \centering    
    \includegraphics[scale=0.93]{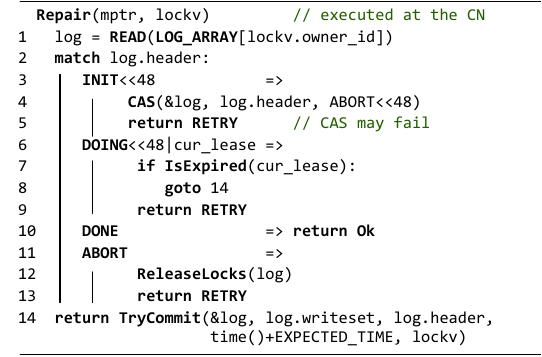}
    \end{minipage} \\[6pt]
    \begin{minipage}{1\linewidth}
    \caption{\small{%
    The simplified pseudo-code of repairing an {\rtx}.
    }}
    \label{fig:recovery}
    \end{minipage} \\[-25pt]
    \end{figure}

{\fig{fig:recovery}} presents the pseudo-code after detecting an expired {\leaselock} with lock state as \texttt{lockv}. 
We first trace the lock owner by reading the log with the help of the ID \texttt{lockv} (Line 1). 
Based on the lock state, our repair actions are follows (Line 2--14):
if the log is \emph{INIT}, 
the {\rtx} has not finished the validation so we just abort it. 
If the log is done, we do nothing.
If the log is aborted, we release its locks. 
{\sys} repairs an {\rtx} if and only if the log is in a doing state with an expired lease. 
To repair, we extract the writeset from the log and call \texttt{TryCommit} (Line 12).

\stitle{Correctness.}
\label{sec:informal-proof}
We informally argue the correctness by showing that 
each {\rtx} behaves as if it was not failed.
First, if an {\rtx} failed and it breaks the failure isolation, 
it must hold some locks. 
So a later affected {\rtx} can detect the failure and resume the execution with our repair phase,
hence the failure is fixed. 
Second, the re-execution of \texttt{TryCommit} will not affect the correctness of other {\rtx}s,
i.e., incorrectly releases their locks or overwrites their values. 
Based on our assumption (unique ID), each {\rtx}'s lock state is unique, 
so the re-execution cannot changed their value.
For data updates it is similar.
Finally, even if the {\rtx} is repaired multiple times,
it is benign because the commit executed by the repair is idempotent.

\subsection{Resource reusing, allocation and reclamation}
\label{sec:resource}

To simplify descriptions,
we assume each {\rtx} will have a unique ID and log buffer
and don't reuse old data pointers in the previous section.
To make {\rtx} practical, we now describe how we reuse these resources.


\begin{figure}[h]
    \begin{minipage}{1\linewidth}
    \hspace{-3mm}
    \centering    
    \includegraphics[scale=1.1]{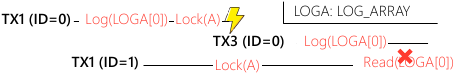}
    \end{minipage} 

\end{figure}

\stitle{Reuse ID and log buffer.}
{\sys} shares the same ID and log buffer for all the {\rtx}s within a CN thread.
However, 
directly reusing the ID and log buffer is incorrect:
a repair {\rtx} may read a wrong log. 
See the above example: suppose TX1 fails and TX2 trying to repair it. 
TX2 first gets the ID of TX1 (0). 
If TX3 reuses TX1's ID and log buffer and write its log before TX2 reads the TX1's log,
TX2 will use a wrong log to repair.

To prevent reading a wrong log buffer, 
we add a checksum at the log header to detect mismatched logs.
The checksum is the LockState of the {\rtx} that writes the log, i.e., calculated with {\rtx}'s lease and ID. 
By keeping the leases of {\rtx}s with the same ID monotonically increasing, 
we can detect a wrong log by comparing the checksum with the \texttt{lockv}
after reading the log at Line 1 in {\fig{fig:recovery}}.
Note that to prevent overwriting an uncommitted {\rtx}'s log buffer, 
a log without a \emph{DONE} state cannot be reused.

\stitle{Data pointer reuse. }
If an {\rtx} commits, the data pointer it updated (\texttt{dptr} at line 5 in {\fig{fig:occ-trycommit}})
is no longer needed,
so we reuse it for future {\rtx}s.
Unfortunately, directly reusing is incorrect due to the famous ABA problem~\cite{DBLP:conf/ppopp/Ben-DavidBW22}:
Suppose TX1 updates the data $A$ from $P_{a0}$ to $P_{a1}$. 
TX2 reads $A$ at $P_{a0}$. 
If TX3 updates $A$ from $P_{a1}$ to $P_{a0}$ again,
TX2 would miss the conflict from TX3 in our protocol (Line 4 in {\fig{fig:occ}}). 
A naive solution is to record a version number at the metadata
that is incremented every time the data is updated.
However, it is not optimal due to extra update and storage overhead. 
Thus, {\sys} piggybacks the version in the {\leaselock}.
To distinguish the lock from version, 
we reserve 1 lock bit in the lockstate. 

Reusing data pointers can also break the idempotence \texttt{TryCommit} (Line 5),
i.e., an unfinished repair overwrite a future {\rtx} that reuses the same data pointer.
Thus, 
we encode a 16-bit version at each data pointer incremented upon updates.
Wrap around can happen, so we further restrict the time to reuse a pointer 
and stop the repair as soon as it may encounter a wrap around.
It is possible to do so because there is a lower bound time on the wrap around.
Specifically, 
we only reuse a data pointer if its {\rtx} is in a done state. 
Afterward, it is safe to abort a repair if its log is done and it execution time is close to this lower bound.
For each repair, we periodically check whether the log state is \emph{done}
with a check interval far less than the wrap around time. 
If the log has been done (or reused by another {\rtx}), we abort the repair. 

\stitle{Memory allocation, reclamation and leakage prevention.}
Detailed memory allocation and reclamation is an orthogonal topic to us~\cite{racehashing,DBLP:conf/sigmod/WangLS22}, 
e.g., with hardware primitives~\cite{254441,genz, DBLP:conf/asplos/GuoSLHZ22}. 
Nevertheless, we still need to handle memory leakage under application failures, 
either due to unused allocated data or old data pointers gathered during {\rtx} executions. 

{\sys} adopts two techniques to prevent memory leakage.
First, we record the allocation information (e.g., bitmap) at the MN,
which is periodically updated by the CNs in a batched way for better performance.
Under batched updates, the MN information can become stale under CN failures. 
After CN reboots, we first fix the stale metadata before processing. 
At a high level, the fix scans all the entry and determines whether the data is used or not.
To detect whether a data is used, we encode an incarnation on the metadata and data pointer, similar to existing work~\cite{farm,racehashing}.
Second, we record a small reclamation table~\cite{DBLP:conf/usenix/XieW0C19} that stores the reused data pointers.
The table is updated and fixed upon reboots similar to the allocation information.

\subsection{Improving {\rtx} performance for DM index}
\label{sec:opt-impl}

A transactional approach---though simplifies failure atomicity, 
introduces overheads---including logging and general concurrency control---to 
index operations. 
{\sys} leverages characteristics of DM index 
and recent advances in transactional approach for concurrent index to reduce such overheads~\cite{DBLP:conf/eurosys/HermanIHTKLS16}. 

\stitle{Opt1. Reducing logging overhead. }
The logging overhead mainly comes from extra \texttt{WRITE} to update the log content 
and \texttt{CAS}es to transfer the log state. 
Logging is not necessary if an index operation only update one data,
since there is no partial write. 
Observing the fact that it is common for DM index to update one value, 
e.g., in trees, multiple updates only happens when there is splits, 
we design a mechanism to completely avoid logging if the writeset size is one. 

Specifically, for {\rtx} that only modifies one object, 
it will leave no log at the {\leaselock}, i.e., with an invalid ID. 
If another {\rtx} reads an invalid ID, 
it will use the current value for the repair:
it will allocate a new data buffer and filled it with the current object value. 
Afterward, we use \texttt{CAS} to switch the data pointer and release the lock. 
The additional \texttt{CAS} is necessary to prevent the suspected {\rtx} from concurrently doing the commits. 
Note that for the object, locking is needed because the object can still be locked by other {\rtx}s.
Avoid this lock is possible, but requires complex index-specific algorithm to detect possible conflicts~\cite{racehashing}, 
and we leave this to the index developers (see below).

\stitle{Opt2. Delegating concurrency control. }
Inspired by recent works on transactional approach for concurrent indexes~\cite{DBLP:conf/eurosys/HermanIHTKLS16},
we allow developers adding customized callbacks to the OCC execution
to avoid unnecessary aborts caused by locking or validation.
For example, {\race} provides lock-free algorithm for its common path,
which can also ensure failure atomicity. 
Therefore, we fully delegate concurrency control to it.  
For {\sherman}, an index update can pre-lock the updated leaf at the execution phase to prevent future conflicts.
Since our methodology mainly follows existing work~\cite{DBLP:conf/eurosys/HermanIHTKLS16}, 
we omit the detailed callback implementations.

\section{Building DM indexes with {\rtx}}
\label{sec:index}

\noindentstitle{Case study: {\sherman}~\cite{DBLP:conf/sigmod/WangLS22}.}
{\sherman} is the state-of-the-art DM {\tree} that adopts a B-link Tree design~\cite{DBLP:journals/dr/Shasha99d}. 
It only achieves concurrency isolation by 
acquiring locks for each updated nodes in a two-phase locking manner. 
Each tree node is 1\,KB by default.

Building transactional {\sherman} with {\rtx} is quite simple. 
We wrap each tree node in a \texttt{TXObj}, 
rewrite the operation to manipulate the tree nodes with {\rtx},
and remove the original concurrency control code. 
To further improve the performance of {\rtx},
we add a callback for leaf node updates, i.e., before updating a leaf node,
we will first acquire the lock instead of delay it to the commit phase. 
This can avoid the aborts for concurrently updating values in a single node.

\stitle{Case study: {\race}~\cite{racehashing}.}
{\race} is a hash index designed for DM. 
It adopts a hierarchical design, 
where the hash buckets are grouped into subtables,
and subtables are managed by a global directory. 
{\race} carefully optimizes its common path operations (e.g., insertion) 
such that they are transactional: 
it only needs a single \texttt{CAS} to update or insert a key-value pair.
On the other hand, if a rehashing happens, 
which means moving part of the keys within a subtable to a new subtable upon full, 
is a non-transactional operations: {\race} leverages spinlocks to coordinate concurrent modifying different subtables.

Since most operations in {\race} are transactional,
we only wrap the \texttt{subtable} of {\race} to a \texttt{TXObj} 
to support transactional rehashing. 
Specifically, if rehashing happens, 
we will use {\rtx} to update the old subtable and insert a new one into the directory.
For insert that only update one buckets in a single table, 
we delegate it to the original {\race}'s algorithm for better performance.
To coordinate the original {\race} algorithm with {\rtx}, 
we add simple checks to the {\leaselock} on the subtable to detect conflicting rehashing 
(including failed rehashing).
This check is lightweight (one \texttt{READ}) and has a small overhead to the performance.

Surprisingly, rehashing with {\rtx} 
is orders of magnitude faster than the original {\race}'s method (milliseconds vs. seconds), 
because {\rtx} updates a table as whole with one transactional write, 
while {\race} updates the table with millions of \texttt{CAS}es. 
The trade-off is that the normal operation may be temporarily blocked due to concurrent rehashing.
We believe it is a reasonable trade-off given the fast rehashing speed and 
the ability to tolerate client failures enabled by {\rtx}.

\section{Evaluation} 
\label{sec:eval}
\label{sec:eval-setup}

We implemented the core {\sys} library with 
10,748 LoC of rust code, excluding tests, benchmarks, indexes 
and RDMA support code. 
Our RDMA library~\cite{krcore} is highly tuned that incorporates 
all known optimizations including huge page~\cite{farm}, coroutines~\cite{drtmh, fasst} and others~\cite{DBLP:conf/sigmod/WangLS22, DBLP:conf/usenix/KaliaKA16, herd}.

\stitle{Experimental setup.} 
Our testbed is a local RDMA-capable cluster with 6 machines:
Each machine has two 12-core Intel Xeon E5-2650 v4 processors and 128\,GB DRAM. 
These machines are bridged via two 100\,Gbps ConnectX-4 MCX455A InfiniBand RDMA-capable NICs 
and one Mellanox SB7890 100Gbps switch. 
Like existing indexes~\cite{racehashing, DBLP:conf/sigmod/WangLS22}, 
we emulate the MN by dedicating a server and running no applications on it.
Other servers are left as CNs.
All CN clocks are loosely synchronized via NTP~\cite{ntp}.

\begin{table}[t]
\vspace{3mm}
\begin{minipage}{.48\textwidth}
    \caption{\small{
    Descriptions of the YCSB benchmarks. 
    \textbf{R}, \textbf{U}, \textbf{I} and \textbf{S} 
    stand for read, update, insert, and scan, respectively. 
    Scan accesses \texttt{N} values, where \texttt{N} is uniformly 
    distributed in [1,100].    
    }}
    \label{tab:ycsb}
    \end{minipage} 
    \small{
        \ra{1.2}
\centering
\begin{tabular}{l|ccccc}
\hline
\textbf{YCSB} & \textbf{A}  & \textbf{B} & \textbf{C} & \textbf{D} & \textbf{E}  \\ 
\hline
{Pattern}        &  {R~:~U}    &  {R~:~U}   & {R}        & {R~:~I}    & {S~:~I}      \\
{Ratio (\%)}  &  {50~:~50}  &  {90~:~10} & {100}      & {95~:~5~}  & {95~:~5~}    \\ 
\hline
\end{tabular} \\[-5pt]

    }
\end{table}

\stitle{Workloads. }
Without explicit notation, 
we use YCSB~\cite{ycsb} to evaluate the performance of different indexes.
YCSB contains different index workloads from real workloads, 
whose patterns are summarized in Table~\ref{tab:ycsb}. 
We load 100 millions key-value pairs to the index, 
where each pair has a payload of an 8\,B key and an 8\,B value,
similar to existing DM indexes~\cite{DBLP:conf/sigmod/WangLS22}.
We evaluate both the uniform and skewed\footnote{\footnotesize{D 
reads the latest inserts instead of following a Zipfan distribution~\cite{ycsb}.}} 
(the request follows a Zipfan distribution with $\theta=0.99$) workloads. 

\stitle{Comparing targets.} 
We compare {\sys} with various implementations of the two state-of-the-art DM indexes, 
{\sherman} and {\race} (see \textsection{\ref{sec:index}}). 
The specific implementations of these indexes are:  \\[-18pt]
\begin{enumerate}[leftmargin=*]
    \itemsep0.5em    
    \setlength{\itemindent}{0em} 
    \item \textbf{\emph{Optimal}} is the original index with no transactional support.
    To enable an apple-to-apple comparison, we reproduced {\sherman} and {\race} with our codebase.
    We enabled all their optimizations and tuned their parameters carefully. 
    Our {\sherman} result matches their open-source code\footnote{\footnotesize{https://github.com/thustorage/Sherman}}
    and {\race} matches their paper's results\footnote{\footnotesize{Not open-sourced.}}.
    These results are further confirmed by their authors. \\[-18pt]

    \item \textbf{\emph{\rtx}} is the index built with {\rtx}. 
    Based on our hardware setup and empirical results (\textsection{\ref{sec:eval-lease}}), 
    we set the estimated time of each transactional update and lease drift to 
    100$\mu$s and 8\,ms (see \textsection{\ref{sec:eval-lease}}), respectively. \\[-18pt]
    
    \item \textbf{\emph{DTX} } is the indexes built with {\drtmh}~\cite{drtmh}---a state-of-the-art DM-based distributed database.
    {\drtmh} adopts a similar protocol as {\rtx} for normal operations,
    except that it adopts a compact design that stores the metadata and data together, 
    using value logging (instead of just recording the pointers like \rtx),
    does not provide idempotent commit and leverage Zookeeper for failure detection and recovery.
    We disabled the replication but enables logging to tolerate client failures.
    The index is built similar to {\rtx} described in \textsection{\ref{sec:index}} as 
    they all provide a transactional interface. 
    We should mention that DTX cannot be directly used for loosely coupled clients and we 
    just use it as a reference solution. \\[-18pt]
\end{enumerate}

\begin{table}[t]
    \vspace{3mm}
    \begin{minipage}{.5\textwidth}
        \caption{\small{
        A summary of the DM operations required to insert an 8\,B value to various {\sherman} implementations,
        assuming all internal nodes are cached at the CN.
        $n$: number of updated nodes. In the common case $n=1$. 
        Note that we omit log header for simplicity.
        }}
        \label{tab:overhead}
        \end{minipage}        
        \ra{1.2}
        \resizebox{.5\textwidth}{!}{%
        \hspace{-3mm}
    \small{            
    \centering
        \begin{tabular}{l|ccc}
            \hline
            & \textbf{READ (Bytes)}            & \textbf{WRITE (Bytes)}               & \textbf{\#CAS (Bytes)} \\ \hline
    \textbf{Optimal} & $1K \times n$                  & $1K \times n + 8$          & $8 \times n$       \\
    \textbf{DTX}     & $(1K + \red{\uline{16}})\times n$            & $\red{\uline{2K}} \times n + 8$          & $8 \times n$       \\
    \textbf{\rtx}     & $(1K + \red{\uline{16}})\times n$            & \begin{tabular}[c]{@{}c@{}}$1K \times n$  + \\ $ \red{\uline{n = 1?}\red{0:n \times 16}} + 8$\end{tabular}  &$\red{\uline{3 \times 8}} \times n$  \\
            \bottomrule
    \end{tabular} 
        }

        }
\end{table}

\noindent
Table~\ref{tab:overhead} summarizes the DM primitives for a {\sherman} insertion, 
the worst case of using {\rtx}.
The detailed overhead analysis on {\sherman} (marked with underlines) 
is in \textsection{\ref{sec:factor-analysis}}.
For {\race}, since its common path operation is mostly executed in its original algorithm, 
we focus on evaluating its rehashing performance in \textsection{\ref{sec:eval-failure}}.

\subsection{Overall performance}
\label{sec:ycsb-overall}

\begin{figure}[!t]
        \hspace{-4pt}
        \includegraphics[left, scale=1.2]{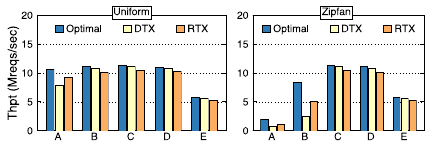}
        \begin{minipage}{1\linewidth}
            \caption{\small{
            Comparison of throughput on various {\sherman} (\tree) implementations.
            }}    \label{fig:ycsb-tree}
        \end{minipage} \\[-20pt]
    \end{figure}
    
    \begin{figure}[!t]
            \hspace{-4pt}
            \includegraphics[left, scale=1.2]{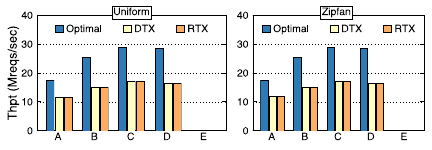}
            \begin{minipage}{1\linewidth}
                \caption{\small{
                Comparison of throughput on various {\race} implementations.
                }}    \label{fig:ycsb-hash}
            \end{minipage} \\[-20pt]
        \end{figure}

{\fig{fig:ycsb-tree}} and {\ref{fig:ycsb-hash}} compares the peak throughput of various implementations of DM indexes. 
We saturate the MN with up to 5 CNs,
each running 24 threads to utilize all the CPU cores.
Note that {\race} doesn't support range queries so we don't evaluate YCSB E on it.
We focus on comparing with {\sherman}'s results.
As shown in {\fig{fig:ycsb-hash}}, 
{\rtx} has a constant performance overhead (two \texttt{READ}s) to check the lease 
in additional to {\race}'s original algorithm.
These operations will degrade the overall throughput as 
{\race} is bottlenecked by small requests.


\stitle{Read-only workloads (YCSB C and E). } 
On {\sherman}, 
{\rtx} is close to the optimal in read-only workloads:
its throughput is 92\% to the original non transactional index
under uniform and skewed workload for YCSB C and E, respectively.
Under read-only workloads, 
{\sherman}'s performance is dominated by network bandwidth (reading 1\,KB tree nodes).
Consequently, the extra overhead of {\rtx}, 
including an additional \texttt{READ} due to the separated metadata layout 
and extra validation overhead (\texttt{READ} another 16\,B) is negligible.
{\rtx} is slightly slower than \textbf{DTX} (7\%) in throughput 
because it issues extra \texttt{READ}s to read metadata. 

\stitle{Read-write workloads (YCSB A, B and D). }
{\rtx} is 87\%, 90\% and 92\% compared to the optimal on the uniform workload for YCSB A, B and D, respectively. 
The overheads mainly come from updates,  which is caused by \texttt{CAS} to update the value and 
extra \texttt{READ}s to fetch the data due to the separated layout. 
For A and B, there is no logging overhead because they only modify one object. 
For D, the logging overhead is negligible because they are rare (one split per 62 insertions).
\textsection{\ref{sec:factor-analysis}} analyzes these overheads in detail. 

On the other hand, 
{\rtx} cannot sustain such a high throughput if the workload is skewed:
it only achieves 58\% and 60\% on YCSB A and B, respectively. 
The performance degradation is mainly due to extra contention on the NIC:
{\rtx} extensively leverage \texttt{CAS} to update values, which will block reads 
due to the NIC's internal locking mechanism~\cite{DBLP:conf/usenix/KaliaKA16}
since the requests focus on a few hot keys.
Compared to \textbf{DTX}, {\rtx} is 1.2--2$\times$ faster on YCSB A and B (except for the uniform case),
thanks to the reduced transaction aborts due to no logging and pre-locking.
\textbf{DTX} on the other hand, suffer from high aborts due to contentions under a skewed workload. 

In general, {\rtx} works best in read-only workloads because they 
don't have the logging and \texttt{CAS} overheads for transactional operations.

\begin{figure}[!t]
    \hspace{-7pt}
    \includegraphics[left, scale=1.2]{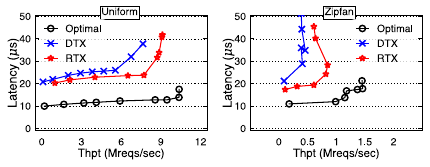} \\[4pt]
    \begin{minipage}{1\linewidth}
    \caption{\small{
        Comparisons of throughput latency of {\sherman} on (a) uniform and (b) skewed
        workloads under YCSB A.
        }}
        \label{fig:latency}
    \end{minipage} 
\end{figure}

\stitle{End-to-end latency. }
{\fig{fig:latency}} further compares the end-to-end latency on {\sherman},
which plots the throughput and latency graph of YCSB A under various workloads. 
We omit other workloads because their trend is similar. 
We run the experiment by increasing the number of clients. 
In general, {\rtx} has a higher latency than others: 
on a single client, 
{\rtx}'s latency is 1.6--2$\times$ higher than the optimal.
This is mainly due to extra round trips to fetch the metadata due to the separate layout
and \texttt{CAS}es for committing.
Though a single \texttt{CAS} has comparable latency as others, 
it may blocks concurrent access to a close memory region on the NIC.
On the other hand, 
{\rtx} is close to \textbf{DTX} even with more roundtrips, 
because it omits the logging for single object update.

\subsection{Handing client failures}
\label{sec:eval-failure}

\begin{figure}[!t]
    \centering
    \includegraphics[scale=1.2]{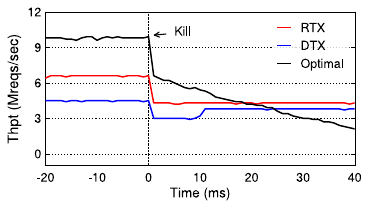} \\[-1pt]
    \begin{minipage}{1\linewidth}
    \caption{\small{
        {\sherman} performance timeline under client failures. 
        }}
        \label{fig:eval-repair}
    \end{minipage} \\[-10pt]
\end{figure}

\noindentstitle{Tolearting client failures in {\sherman}.}
To evaluate the performance with failures, 
we ran a microbenchmark where all clients inserting values to {\sherman},
and we kill the benchmark client on a CN after all clients warm up. 
The throughput of different clients are aggregated at 1\,ms intervals with RDMA messaging, 
similar to priori work~\cite{dragojevic2015nocompromises}.

{\fig{fig:eval-repair}} shows the timeline under client failures. 
When the client is killed, 
{\rtx} has no observable performance collapse except for the degraded overall performance (6.5 vs. 4.2 Mreqs/sec),
which is due to the reduced number of clients (from three to two).
{\rtx} has such a fast recovery thanks to our design of on-demand, fine-grained index repair, 
where each individual repair (recommit an operation) is finished in a few microseconds
without coordinating with other clients.
In comparison,
\textbf{DTX} has a performance drop because it needs a complicated 
protocol to coordinate different clients~\cite{dragojevic2015nocompromises,DBLP:conf/eurosys/ChenWSCC16}.
Note that we configure DTX with the same failure detection interval (8\,ms) as leases in {\rtx}.
Finally, \textbf{Optimal}'s throughput degrades gradually because 
more clients are being blocked by failed client's locks.

\begin{figure}[!t]
    \vspace{-5pt}
    \centering
    \includegraphics[scale=1.2]{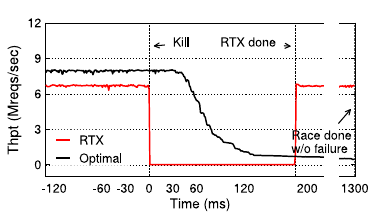} \\[-1pt]
    \begin{minipage}{1\linewidth}
    \caption{\small{
        {\race} insertion timeline under failed rehashing. 
        We omit DTX as it's similar to {\rtx}.
        }}
        \label{fig:eval-repair-rehash}
    \end{minipage} \\[-10pt]
\end{figure}

\stitle{Tolearting client failures under {\race} rehashing. }
Next we present the timeline of different {\race} implementations under client failures
in {\fig{fig:eval-repair-rehash}}. 
Three CNs  spawn multiple threads concurrently inserting to the index.
Like {\sherman}, if we kill a client that is rehashing,
we will see a gradual throughput degradation in the original {\race} (\textbf{Optimal}),
because clients are blocked waiting for the rehashing to be done.
It is impossible in the original {\race} to tolerate client failures because 
the client will hold the locks on the subtables for rehashing. 
On the other hand, 
{\rtx} can recover failure in 185\,ms, thanks to the transactional property ensured by it. 
More importantly, its rehash speed is orders of magnitude (200\,ms vs. 1.3\,secs) faster than 
the original {\race} (considering not failed). 
It only needs 2 \texttt{CAS}es to install the changes to the hash table.
In comparison, the original {\race} needs scan all the buckets and issue 2
\texttt{CAS}es for each KV-pair migrated (more than 500 millions pairs moved in average).

\subsection{Factor analysis}
\label{sec:factor-analysis}

\begin{figure}[!t]
    \centering
    \includegraphics[left, scale=1.2]{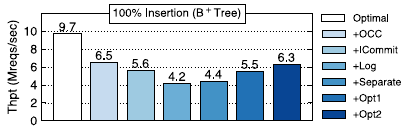}
    \begin{minipage}{1\linewidth}
        \caption{\small{
        Analysis of {\rtx} overhead to the peak throughput.
        }}    \label{fig:e2e-factor-analysis}
    \end{minipage} 
\end{figure}

\begin{figure}[!t]
    \centering
    \includegraphics[left, scale=1.2]{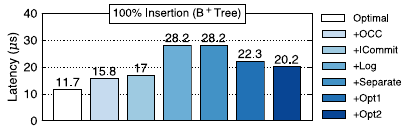}
    \begin{minipage}{1\linewidth}
        \caption{\small{
        Analysis of {\rtx}} overhead to average latency.
        }    \label{fig:e2e-factor-analysis-lat}
    \end{minipage} \\[-10pt]
\end{figure}

To analysis the overhead imposed by {\rtx} as well as our optimizations for DM indexes, 
we conduct a factor analysis on the {\sherman}
by breaking down the performance gap between \textbf{Optimal} and {\rtx}.
We choose a workload with 100\% insertion, 
which reflects all the possible overhead introduced by {\rtx}.
{\fig{fig:e2e-factor-analysis}} and {\fig{fig:e2e-factor-analysis-lat}} presents the results 
on the throughput and latency, respectively.
We measure the latency a single CN to prevent interferences from queuing effect. 

\stitle{+OCC. }
A general concurrency control protocol like OCC incurs 33\% throughput degradations 
and 1.35$\times$ latency increases, 
due to extra round trip for readset validation
 (the extra 16\,B \texttt{READ} in the Table~{\ref{tab:overhead}}).

\stitle{+ICommit. }
Turning the commit phase in OCC to realize idempotence further adds 14\% throughput degradations
(the two 8\,B \texttt{CAS}),
because it leverages \texttt{CAS} to update the values and releases locks.
Compared to the normal update implemented with \texttt{WRITE}, 
the atomic operations is more costly to execute on the NIC 
due to extra read, NIC internal locking and computation.
The latency impact is limited because \texttt{CAS} has the same execution time as \texttt{WRITE} 
when the NIC is not saturated. 

\stitle{+Log.}
Add write-ahead logging of \textbf{DTX} for tolerating non-atomic updates 
degrades the throughput by 26\% and increases the latency by 1.7$\times$.
Each log contains the entire internal nodes modified (2\,KB), which is non trivial.

\stitle{+Separate. }
Compared to \textbf{DTX}, {\rtx} further separate the object data with metadata, 
reducing the log payload from 1\,KB to 16\,B for each node. 
The trade-off is that read wil introduce extra roundtrip to read the metadata 
(the overhead is not shown in Table~\ref{tab:overhead}).
It is acceptable: 
the throughput is improved by 1.06$\times$ due to reduced log size, 
while the latency is remain unchanged. 

\stitle{+OPT1 +OPT2.} 
Finally, our optimizations \textsection{\ref{sec:opt-impl}} 
improves the throughput by 1.43$\times$ and reduce the latency by 
30\%, respectively.
\textbf{OPT1} removes most of the logging overhead in {\rtx} ($n=1$ case), 
since tree insertion happens infrequently. 
\textbf{OPT2} further reduces the validation and abort costs of OCC.

\subsection{Impact of the choice of lease time}
\label{sec:eval-lease}

\begin{figure}[!t]
    \hspace{-5pt}
    \includegraphics[left, scale=1.2]{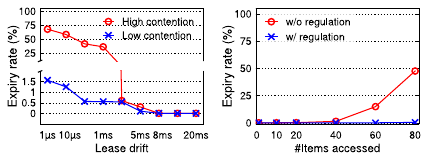} \\[4pt]
    \begin{minipage}{1\linewidth}
    \caption{\small{
        (a) False expiry rate of lease locks with different drifts and 
        (b) how the number of the locked items imapact false expiry.
        }}
        \label{fig:eval-lease}
    \end{minipage} 
\end{figure}

To evaluate the accuracy of lease, 
we ran an experiment where all CN threads are alive and count the lease expiry rate.
Each {\rtx} updates various numbers of objects.

\stitle{Single-lock (false) expiry rate. }
Each {\rtx} only update one object. 
Since the expiry rate depends on how frequently {\rtx}s access the same object, 
we ran two setups:
in the high contention setup, 72 clients update 100 objects. 
In the low contention setup, they modifies 10,000 objects. 
The results in {\fig{fig:eval-lease}} (a) show that in a loosely synchronized cluster, 
a lease drift of 8\,ms is sufficient to eliminate most false expiry, 
even under a high contention workload. 
Specifically, we met no false expiry for drift larger than 8\,ms 
and the false expiry rate is less than 1\% for a 5\,ms drift.

\stitle{Multi-locks (false) expiry rate. }
The commit time may also impact the expiry rate, 
so {\sys} regulates the lease based on the number of items read amd written by it. 
Our results in {\fig{fig:eval-lease}} (b) show that regulation is a necessary step:
with a fixed lease using an 8\,ms drift (\evalline{w/o regulation})
the false expiry rate increases from 0.1\,\% to 47.5\,\% when {\rtx} updates from 20 to 80 objects.  
With regulation, only updating 80 items meet a 0.4\,\% false expiry. 
\section{Discussion}
\label{sec:dislim}

\noindentstitle{Performance overhead and trend. }
The overhead of {\rtx} mainly comes from the following aspects:
(1) the extra round trip caused by a meta-data separate design, 
which is used for pointer chasing,
(2) the extra \texttt{CAS} used to update the data record in an idempotent way and 
(3) logging overhead for failure atomicity.
While the logging overhead is unavoidable for {\rtx} with more than one updates, 
the overhead of (2) is acceptable thanks to recent improvements in NIC hardware for faster \texttt{CAS}~\cite{drtmh,DBLP:conf/usenix/KaliaKA16}.
Meanwhile, 
we see a trend in hardware improvements for reducing the overhead of pointer chasing.
Specifically, there has been proposals and implementations 
to add a pointer chasing primitive to DM interconnects or hardware~\cite{DBLP:conf/iccd/HsiehKVCBGM16,DBLP:conf/fpga/WeiszMWFNH16,DBLP:conf/hotos/AguileraKNS19, DBLP:conf/sosp/0001DJSNZP21},
e.g., with indirect RDMA \texttt{READ}~\cite{DBLP:conf/sosp/0001DJSNZP21}.
With the help of it, {\rtx} can use one DM operation (instead of two) to read an object.

\stitle{Tolerating memory node failures. }
Though DM index applications like caching don't need tolerating MN failures,
{\sys} may support \emph{durability} by 
placing the data on the persistent memory, like previous work~\cite{ford-fast22,DBLP:conf/sosp/AguileraMSVK07},
or periodically checkpointing the in-memory data to other persistent storage.
On the other hand, 
supporting \emph{high availability} requires replicating the data on multiple memory nodes 
and coordinating clients' accesses to them with a global coordination service~\cite{fuse-fast23,dragojevic2015nocompromises,redis-sentinel}, which is out of the scope of this paper. 

\section{Related Work}
\label{sec:related}

\noindentstitle{DM or RDMA indexes. }
To the best of our knowledge, no existing DM indexes~\cite{racehashing,DBLP:conf/sigmod/WangLS22, DBLP:conf/hotos/AguileraKNS19, DBLP:conf/sigmod/0001VBFK19, rolex} 
provides failure atomicity. 
Though existing RDMA-based indexes~\cite{xstore,cell,mitchell2013pilaf, farm, fasst, DBLP:journals/usenix-login/KaliaKA19, herd} 
provides failure atomic and isolated index updates, 
they all needs RPC for the atomic updates, 
so their techniques are not directly applicable to the setup of DM.

\stitle{Fault-tolerance in DM key-value store. }
A recent work---FUSEE~\cite{fuse-fast23}---also noticed the importance of building fault tolerant 
key-value store atop of DM.
It brings part of the fault tolerance to {\race} by 
replicating the index on multiple memory nodes 
to tolerate both the MN and CN failures. 
Unfortunately, its technique cannot directly handle the atomicity of rehashing 
since it does not provide a transactional semantic, which 
we have confirmed with their authors. 
{\rtx} rightly handles the client failures even under rehashing with {\rtx},
and we believe our approach can further complement the current design of FUSEE.
Meanwhile, FUSEE also does not support more complex indexes like {\tree}.

\stitle{Other disaggregated memory systems. }
DM has received interest from both academia and industry.
Recent work has built various systems to better support applications under disaggregated memory architecture,
including but not limited to operation system~\cite{DBLP:conf/usenix/ShanHCZ19,DBLP:conf/nsdi/GuLZCS17, DBLP:conf/eurosys/AmaroBLOAPRS20}, 
hardware support~\cite{DBLP:conf/isca/LimCMRRW09,DBLP:conf/hpca/LimTSACRW12}, 
memory management~\cite{remote-region, DBLP:conf/osdi/RuanSAB20, DBLP:conf/usenix/TsaiSZ20,DBLP:conf/osdi/WangMLLRNBNKX20} 
and others~\cite{10.1145/3152434.3152447,DBLP:conf/sigcomm/CostaBRK15,DBLP:conf/nsdi/ShrivastavVBCLW19, DBLP:conf/osdi/GaoNKCH0RS16}.
We focus on tolerating client failures in DM indexes,
but we believe similar problem may exist in other DM systems. 
For example, to support Linux, LegoOS~\cite{DBLP:conf/usenix/ShanHCZ19} may share Linux's kernel states between pComponents (clients) on the same MN, 
so failed pComponents can break failure isolation between different pComponents.

\stitle{Transactional indexes.}
{\sys} is also inspired by existing studies on transactional memory,
including software transactional memory~\cite{DBLP:conf/eurosys/HermanIHTKLS16,DBLP:conf/pldi/SpiegelmanGK16,DBLP:conf/wdag/DiceSS06} 
and persistent transactional memory~\cite{DBLP:conf/asplos/LiuZCQWZR17,DBLP:conf/asplos/RamanathanKMFDM20,DBLP:conf/usenix/GuYWWZGC19},
which aims at building concurrent indexes with transactions.
{\rtx} further considers the client failure isolation property under disaggregated memory,
which is not needed in a single-machine setup of these systems.

\section{Conclusion}
\label{sec:concl}

Tolerating clients (computing nodes) failures is critical yet challenging under disaggregated memory,
especially for DM indexes.
We argue a practical DM index should be transactional:
the index operation should be failure atomic and failure isolated in additional 
to concurrency isolated.
We present {\sys}, a systematic approach for building transactional DM indexes 
with a lightweight primitive called repairable transactions (\rtx).
{\sys} has shown effective for two representative DM indexes: 
{\race} and {\sherman}.

\balance

\small{
\bibliographystyle{acm}
\bibliography{a}
}

\clearpage

\end{document}